\documentclass[useAMS,referee,usenatbib]{biom}
\def\bSig\mathbf{\Sigma}

\usepackage[figuresright]{rotating}
\usepackage{graphicx}

\title[Towards Automated Animal Density Estimation with Acoustic Spatial Capture-Recapture]{Towards Automated Animal Density Estimation with Acoustic Spatial Capture-Recapture}

\author{Yuheng Wang$^{1,*}$\email{yw99@st-andrews.ac.uk}, 
Juan Ye$^{2,**}$\email{jy31@st-andrews.ac.uk}, and 
David L. Borchers$^{1,***}$\email{dlb@st-andrews.ac.uk} \\
$^{1}$Centre for Research into Ecological and Environmental Modelling,  School of Mathematics \\ and Statistics, University of St Andrews, The Observatory, St Andrews, Fife, KY16 9LZ, \\ Scotland. \\
$^{2}$School of Computer Science, University of St Andrews,  North Haugh, St Andrews, Fife, \\ KY16 9SX, Scotland.}

\begin{document}


\date{{\it Received October} 2023. {\it Revised February} 2023.  {\it
Accepted March} 2023.}



\pagerange{\pageref{firstpage}--\pageref{lastpage}} 
\volume{64}
\pubyear{2023}
\artmonth{November}


\doi{nothing!}


\label{firstpage}


\begin{abstract}

Passive acoustic monitoring can be an effective way of monitoring wildlife populations that are acoustically active but difficult to survey visually. Digital recorders allow surveyors to gather large volumes of data at low cost, but identifying target species vocalisations in these data is non-trivial. Machine learning (ML) methods are often used to do the identification. They can process large volumes of data quickly, but they do not detect all vocalisations and they do generate some false positives (vocalisations that are not from the target species). Existing wildlife abundance survey methods have been designed specifically to deal with the first of these mistakes, but current methods of dealing with false positives are not well-developed. They do not take account of features of individual vocalisations, some of which are more likely to be false positives
than others. We propose three methods for acoustic spatial capture-recapture inference that integrate individual-level measures of confidence from ML vocalisation identification into the likelihood and hence integrate ML uncertainty into inference. The methods include a mixture model in which species identity is a latent variable. We test the methods by simulation and find that in a scenario based on acoustic data from Hainan gibbons, in which ignoring false positives results in 17\% positive bias, our methods give negligible bias and coverage probabilities that are close to the nominal 95\% level.

\end{abstract}

%



\maketitle


%

\section{Introduction}
\label{s:intro}

Acoustic surveys can be an effective means of assessing wildlife populations that are vocally active but difficult to see. The use of passive acoustic monitoring methods is advancing rapidly, as it causes less disruption and impact on target species than physical traps. There is a variety of spatial capture-recapture (SCR) methods that use an array of acoustic detectors to survey acoustically active species and identify which detections on different detectors are of the same vocalisation \citep[These constitute the ``recaptures''; see][for a review of SCR]{10.2307/24780847} for animal density estimation. However, identifying calls manually in the recordings is labour-intensive and time-consuming \citep{Somervuo2006}, when acoustic detection is by means of digital recorders deployed in the field for long periods. 

Machine learning (ML) methods provide an effective option for automated call identification; e.g. birds \citep{Cakir2017_8081508}, marine mammals \citep{Jiang2019}, amphibians \citep{LeBien2020}. These methods have achieved promising detection accuracy, which makes long-term, large-scale acoustic surveys feasible. 
The ML detection process does make errors, both missing some target species calls (false negatives) and incorrectly identifying other sounds as target species calls (false positives). Statistical methods for wildlife surveys are designed to deal with false negatives although the detection functions used to do this are different for automated detectors and human detectors. While methods have been developed for dealing with false positives, these are in the form of correction factors applied after applying statistical methods that assume no false positives. Methods that deal with false positives explicitly within the statistical model used for inference remain to be developed. This is what we do in this paper, for acoustic spatial capture-recapture (ASCR) inference. There are two main issues that we need to address: how to integrate false positives in inference, and how to modify the detection function used in inference to be appropriate when detection is by an ML method instead of by humans. 

While both human and ML identifiers may generate false negatives (i.e., missing some vocalisations of the target species), humans are typically assumed to produce no false positives, whereas ML identifiers invariably do this to a greater or lesser extent. False positive has been studied in distance sampling by \citet{10.1121/1.3089590} \citet{10.1121/1.3583504} and \citet{10.5751/ACE-01224-130207}, mark-recapture~\citep{10.1121/1.3662070}, and SCR~\citep{https://doi.org/10.1111/j.1748-7692.2011.00561.x}. For example, \citet{https://doi.org/10.1111/brv.12001} gives the following canonical estimator of density:
\begin{equation} \label{equ:canonical}
    \hat{D} = \frac{N(1-\hat{f})}{\hat{p}a\hat{r}}
\end{equation}
where $N$ is the number of vocalisations detected, including false positives, $(1-\hat{f})$ is the false positive correction factor, where $\hat{f}$ is an estimate of the false positive rate which is often obtained from a separate dataset, $\hat{p}$ is an estimate of detection probability within the survey region of area $a$, and $\hat{r}$ is an estimate of the expected number of vocalisations per animal. 

This general form is widely used; however, it has several drawbacks. It employs the false positive rate $\hat{f}$ from an independent dataset, which may not be appropriate for the current survey because acoustic recordings from different datasets may have different properties, which result in different false positive rates. In addition, the detection probability $\hat{p}$ is estimated from data that includes false positives and may be biased as a result, because sounds from the target species may have different detectability to other sounds.

Another approach, whose applicability will depend on the nature of the survey process, is to drop all the vocalisations detected by less than two detectors, assuming they are unlikely to be false positives~\citep{https://doi.org/10.48550/arxiv.2207.09343}. At best this discards some information and more generally it may not get rid of all false positives.


Because the probability of detecting the target species using ML methods is different from that for manual detection, we adapt the threshold models of \citet{stevenson2015} and \citet{https://doi.org/10.1890/08-1735.1} that have been used for human detectors, to be more appropriate for the ML context by doing away with detection threshold and instead modelling detection probability as a smooth function of received signal strength. 

In this paper, we develop a robust framework for integrating ML output into inference for ASCR surveys. The novelty of our method lies in the fact that we use the ML measure of confidence that detection is from the target species \citep{pmlr-v70-guo17a} as a covariate, and we treat species identity as a latent variable.

\section{Methods}\label{sec:method}
Our method is made up of three components: (1) developing a detection probability model that is a smooth function of received signal strength, (2) dealing with false positives using the ML output confidence measure as a covariate, and (3) a bootstrap procedure for interval estimation.

\subsection{Notation and Terminology}
We consider a survey with a duration $T$ in a survey region $a \subset {\rm I\!R}^2$  using $M$ microphones placed at known locations in $a$. An ML technique will be employed to detect calls of target species on audio recordings from each microphone collected during the survey period. We assume that the same call can be detected by more than one microphone. After the ML detection, the observation data consists of $N$ unique detected calls, with a capture history $\bm{\Omega}$, received signal strengths $\bm{Y}$, times of arrival $\bm{Z}$ measured from the beginning of the survey, and detection confidence outputs $\bm{P}$ from the ML technique. 

More specifically, a binary capture history $\omega_{n,m}$ is 1 if the call $n \in \{1, ..., N\}$ is detected at the microphone $m \in \{1, ..., M\}$, and 0, otherwise. The capture history for the call $n$ across all the $M$ detectors is denoted $\bm{\omega}_n = (\omega_{n,1}, ..., \omega_{n,M})$, while $\bm{\Omega} = (\bm{\omega}_1, ..., \bm{\omega}_N)$ is the capture history for all the calls. $y_{n, m}$ and $z_{n, m}$ are the signal strength and  recording time of call $n$ detected at microphone $m$. The ML output is a measure of confidence that a call $n$ detected on microphone $m$ is from the target species, and we denote this $\rho_{n, m}$. We have denote elements $(n,m)$ of the matrices $\bm{Y}$, $\bm{Z}$, and $\bm{P}$ as $y_{n,m}$, $z_{n,m}$, and $\rho_{n,m}$, respectively.

Because some of the calls identified by the ML process may not be from the target species, we define latent variables $\bm{\zeta} = (\zeta_{1}, ..., \zeta_{N})$ such that $\zeta_n = 1$ if a call $n$ is from the target species (a true positive), and $\zeta_n = 0$ if it is not (a false positive). The detected calls come form unobserved (``latent'') locations given by Cartesian coordinates $\bm{X} = (\bm{x}_1, ..., \bm{x}_N)$.

In the following, for brevity and readability, we do not usually show parameters as explicit arguments of the functions we develop.

\subsection{Automated ASCR Model without False Positives}\label{sec:automatic_ascr}
In this section, we first describe the ML continuous detection probability model and how it fits into the ASCR likelihood function without assuming the existence of false positives. 

\subsubsection{Call Detection Function}
Detection probability from ML depends on received signal strength but ASRC requires detection probability to be parameterised as a function of the distance $d$ of the sound source from the detector. We model the binary detection indicator $\omega$ conditional on $d$ and $\zeta$ as a Bernoulli random variable with probability density function (pdf) $f(\omega | d, \zeta)$ with the Benroulli parameter $g(d, \zeta) = p(\omega = 1 | d, \zeta)$ being the probability of detecting a call, given the target species indicator $\zeta$ and the distance $d$ of the call from the detector. 

\begin{figure}[]   
     \centering
     \includegraphics[width=6.5in]{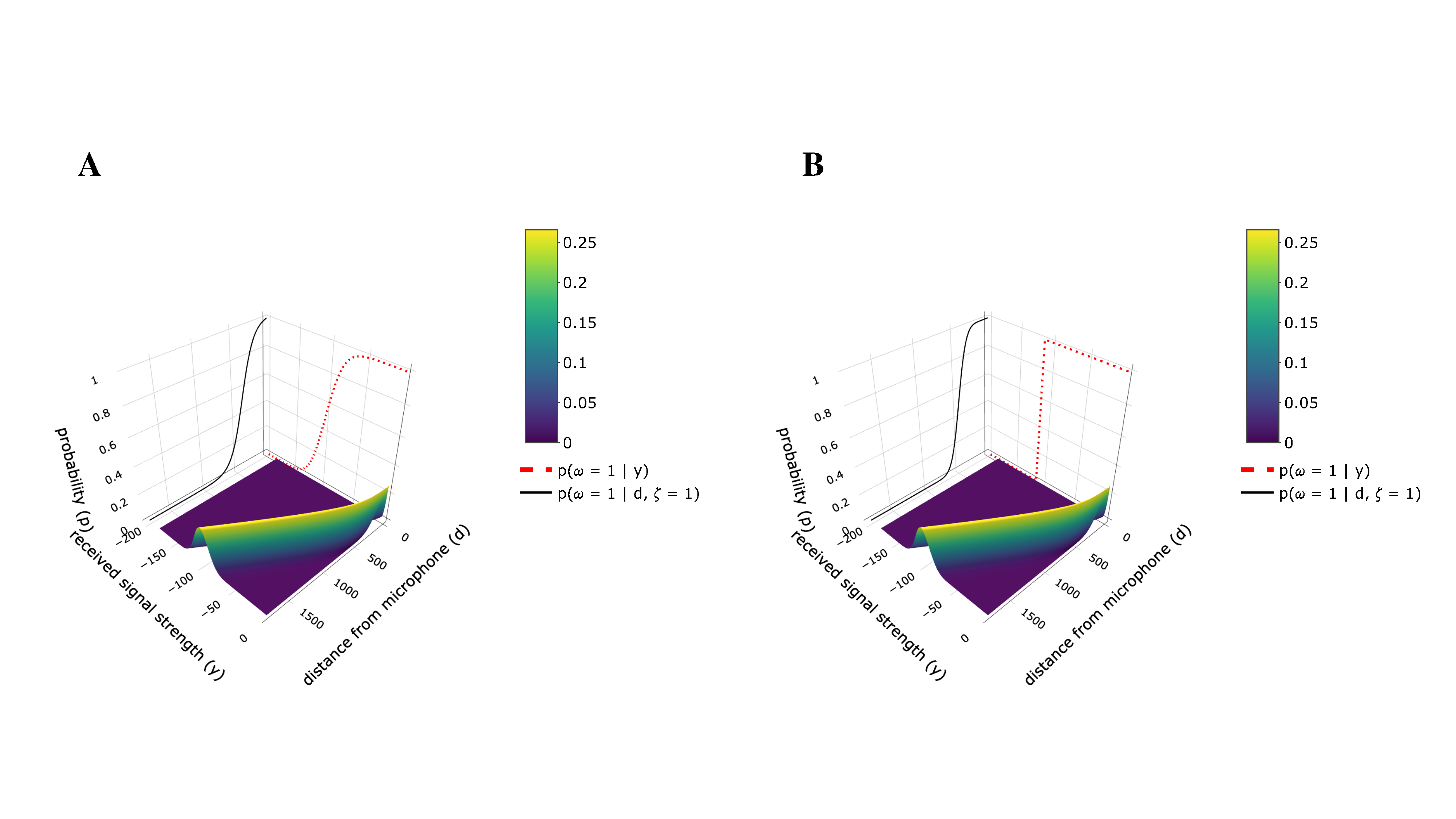}
     \caption{The detection function and its components. In each plot, the base shows the distribution of received signal strength of the target species as a function of source strength and distance $d$ from the microphone, $f(y | d, \zeta = 1)$. The dashed curve is the probability of detecting a call given received signal strength, $p(\omega = 1 | y)$, and the solid curve is the probability of detecting a call given the distance, $p(\omega = 1 | d, \zeta)$.  
     Panel \textbf{A} shows the form of $p(\omega = 1 | y)$ and the corresponding $p(\omega = 1 | d, \zeta)$ that we use, while Panel \textbf{B} shows these when using the step function form for $p(\omega = 1 | y)$ that has been used by other authors. This figure appears in color in the electronic version of this article.
     }
     \label{fig:compare_auto_step}
\end{figure}

Like \citet{stevenson2015} and \citet{https://doi.org/10.1890/08-1735.1}, we construct the distance-dependent detection function $g(d, \zeta)$ by modelling the distribution of received signal strength $y$ at a microphone as a random variable whose mean depends on distance $d$, and a model for the probability of detection as a function of received signal strength. 
Figure~\ref{fig:compare_auto_step} illustrates this.

In Figure \ref{fig:compare_auto_step}, the ``hill'' on the base represents the pdf of received signal strength, in which a slice through the hill perpendicular to the distance axis at a distance $d$ is the pdf of received signal strength from the target species at a distance $d$ from a microphone. Following \citet{https://doi.org/10.1890/08-1735.1} and \citet{stevenson2015}, we assume that the pdf of received signal strength $y$ for true positives ($\zeta=1$) is a Normal distribution with variance $\sigma^2_s$ and a mean that is a monotonically declining function of the distance $d$ between call and microphone, with parameters  $\beta_0, \beta_1$ (see Web Appendix A for details). We denote this pdf $f(y | d, \zeta = 1)$:
\begin{equation}
     f(y | d, \zeta = 1) = N(y| E[y|d; \beta_0, \beta_1], \sigma_{s}^2)
\end{equation}

The dashed curves on the back right panel in Figure \ref{fig:compare_auto_step}\textbf{A} show the probability of detecting a call from the target species, $p(\omega = 1 | y)$, which we assume to depend on received signal strength only, and not on $\zeta$. The step function in Figure~\ref{fig:compare_auto_step}\textbf{B} is the functional form used by \citet{stevenson2015} and \citet{https://doi.org/10.1890/08-1735.1}, with the step occurring at a threshold signal strength value above which calls are certain to be detected and below which they are not detected or are discarded. In contrast, the continuous function on Figure~\ref{fig:compare_auto_step}\textbf{A} shows the form of the signal strength-dependent detection function that we use with automated call detection. Instead of a threshold, it assumes a smoothly increasing probability of detection as received signal strength increases. Here we assume that $p(\omega =1 | y)$ has logistic functional form: $p(\omega = 1 | y) = \left[ 1+ e^{-(r_0 + r_1 y)}\right]^{-1}$, where $r_0$ and $r_1$ are parameters to be estimated. The form is determined by the ML model by testing on the labelled dataset (see Web Appendix B for details). 

The smooth solid curve on the back left panel of Figure \ref{fig:compare_auto_step} shows the resulting distance-dependent detection function for the target species, $g(d, 1)$.  It is obtained by taking the product of the signal strength-dependent detection function and the signal strength pdf and integrating out the received signal strength:
\begin{equation} \label{equ:dp_auto}
    g(d, 1) = \int_{-\infty}^{\infty}p(\omega =1 | y)f(y | d, \zeta = 1)dy
\end{equation}
In our implementation, we use an approximation of this integral to speed up the evaluation of the associated likelihood function (see Appendix 
A for details).

\subsubsection{Likelihood for Automated ASCR without False Positives}

Similar to \citet{stevenson2015}, the likelihood that we use for point estimation assumes that the call locations of the target species are independent draws from a pdf $f(\bm{x}_n| \zeta_n = 1)$. See Section \ref{sec:bootstrap} below for interval estimation. The contribution from the detected call $n$ to the conditional likelihood function, given detection by at least one microphone (i.e. all the following components are conditioning on the call being detected at least once, we omit this universe condition for simplicity), is obtained as the product of the following four terms and their pdfs are given in Appendix B
:
\begin{enumerate}
\item \textbf{the pdf of received signal strengths} $\bm{y}_n$, given the capture history $\bm{\omega}_n$ and source location $\bm{x}_n$: $f(\bm{y}_{n}|\bm{\omega}_n,\bm{x}_n, \zeta_n =1)$, which depends on parameters $\bm{\gamma}=(r_0$, $r_1$, $\beta_0$, $\beta_1$, $\sigma_s)$;

\item \textbf{the pdf of detection times} $\bm{z}_n$, given the capture history $\bm{\omega}_n$ and source location $\bm{x}_n$: $f(\bm{z}_n|\bm{\omega}_n, \bm{x}_n)$, which depends on a parameter $\bm{\phi} = (\sigma_t)$; 


\item \textbf{the pdf of the capture history} $\bm{\omega}_n$, given source location $\bm{x}_n$: $f(\bm{\omega}_n|\bm{x}_n, \zeta_n = 1)$, which depends on parameters $\bm{\gamma}$; 

\item \textbf{the pdf of the source location} $\bm{x}_n$: $f(\bm{x}_n| \zeta_n = 1)$ which in general depends on some parameter(s) $\boldmath{\theta}$, but here we assume a bivariate uniform distribution. Also, since an observation has to be detected by at least one microphone, the location also depends on detection parameters $\bm{\gamma}$. 
\end{enumerate}

Assuming independent detections, and marginalising over $\bm{x}$, this leads to the conditional (on detection) likelihood, assuming no false positives:
\begin{longequation} 
\begin{array}{ll}%
L_{tp}(\bm{\gamma},\bm{\phi}) 
&= \prod_{n=1}^{N} \int_{A} f(\bm{y}_n, \bm{z}_{n}, \bm{\omega}_n,\bm{x}_n | \zeta_n = 1;\bm{\gamma}, \bm{\phi})d\bm{x} \\
 &=\prod_{n = 1}^{N} \int_{A} 
 f(\bm{y}_{n}|\bm{\omega}_n,\bm{x}_n, \zeta_n =1) 
 f(\bm{z}_n|\bm{\omega}_n, \bm{x}_n)
 f(\bm{\omega}_n|\bm{x}_n, \zeta_n = 1) 
 f(\bm{x}_n| \zeta_n = 1) 
 d\bm{x}
 \end{array}
\label{equ:sascr}
\end{longequation}


Aside from the different form for $p(\omega=1|y)$, this is the same likelihood as that of \citet{stevenson2015}. We estimate the parameters $\bm{\gamma}$ and $\bm{\phi}$ by maximising the log of the above likelihood with respect to these parameters.

\subsubsection{Call Density Estimator}
Given the maximum likelihood estimates of $\bm{\gamma}$ and $\bm{\phi}$, the call density (number of calls per unit area per unit time) can be estimated using a Horvitz Thompson-like estimator:
\begin{equation} \label{equ:HTL_estimator}
    \hat{D_c} = \frac{N}{\hat{p}a T}
\end{equation}
\noindent
where $a$ is the area of the survey region, $T$ is the survey duration, and $\hat{p}$ is the maximum likelihood estimator of the mean detection probability in the survey region. $\hat{p}$ is obtained by evaluating
\begin{equation} \label{equ:p}
    p = \frac{\int_a p.(\bm{x} | \zeta = 1)\;d\bm{x}}{a}
\end{equation}
\noindent
at the maximum likelihood estimators of $\bm{\gamma}$ and $\bm{\phi}$, where $p.(\bm{x} | \zeta = 1)=1 - \prod_{m=1}^{M} 1 - g(d_{m}(\bm{x}), 1)$ is the probability that a call is detected by at least one microphone at given location $\bm{x}$, and $d_{m}(\bm{x})$ is the distance from a call location $\bm{x}$ to the microphone $m$. 

Notice that if we divide $\hat{D_c}$ by an estimate of the mean call rate per individual, $\hat{\mu_c}$, then we get an estimator of the same form as the proposed by \cite{https://doi.org/10.1111/brv.12001} in Eq~(\ref{equ:canonical}), with $\hat{r}=\hat{\mu_c}T$, but without the correction $(1-\hat{f})$ for false positives.

\subsection{Tackling False Positives}\label{sec:deal_fp}
Most (but not all) methods of estimating absolute wildlife abundance are designed to cope with false negatives (e.g. missed calls) but are sensitive to false positives (e.g. using sounds that are not the call from the target type). We propose methods that use the confidence measure output by an ML detection algorithm into the ASCR model while the confidence output is a number between 0 and 1 or a positive real number in $(0, \infty)$, quantifying the confidence that detection is the true positive. 

We propose three models that use the confidence output to deal with false positives. Two involve a mixture of models for true positives and false positives. The \textit{fixed-confidence} mixture model treats the ML confidence output as the mixture weight, while the \textit{random-confidence} mixture model treats the ML confidence output as observations of a random variable. The third model is to use the confidence output as a power term in the likelihood, weighting the likelihood contribution for each observation, which we called the \textit{pseudo-likelihood} model. 

\subsubsection{Detection Function for False Positives}

We assume the same kind of model for the probability of detecting a signal that has been identified by the automated detector as a target species call but is not actually a target species call, as assumed for the true positives above. The difference is that we assume that the received signal strength from these non-target calls arise from a different distance-dependent pdf $f(y | d, \zeta = 0)$ than that for target species calls. The non-target calls may have more than one type of distance-dependent pdf, but we model them using a single mode as our goal is to separate these false positives rather than accurately estimating their detection parameters. Other forms of non-target call pdf might be used within our framework.

\begin{figure}[]   
         \centering
         \includegraphics[width=6.5in]{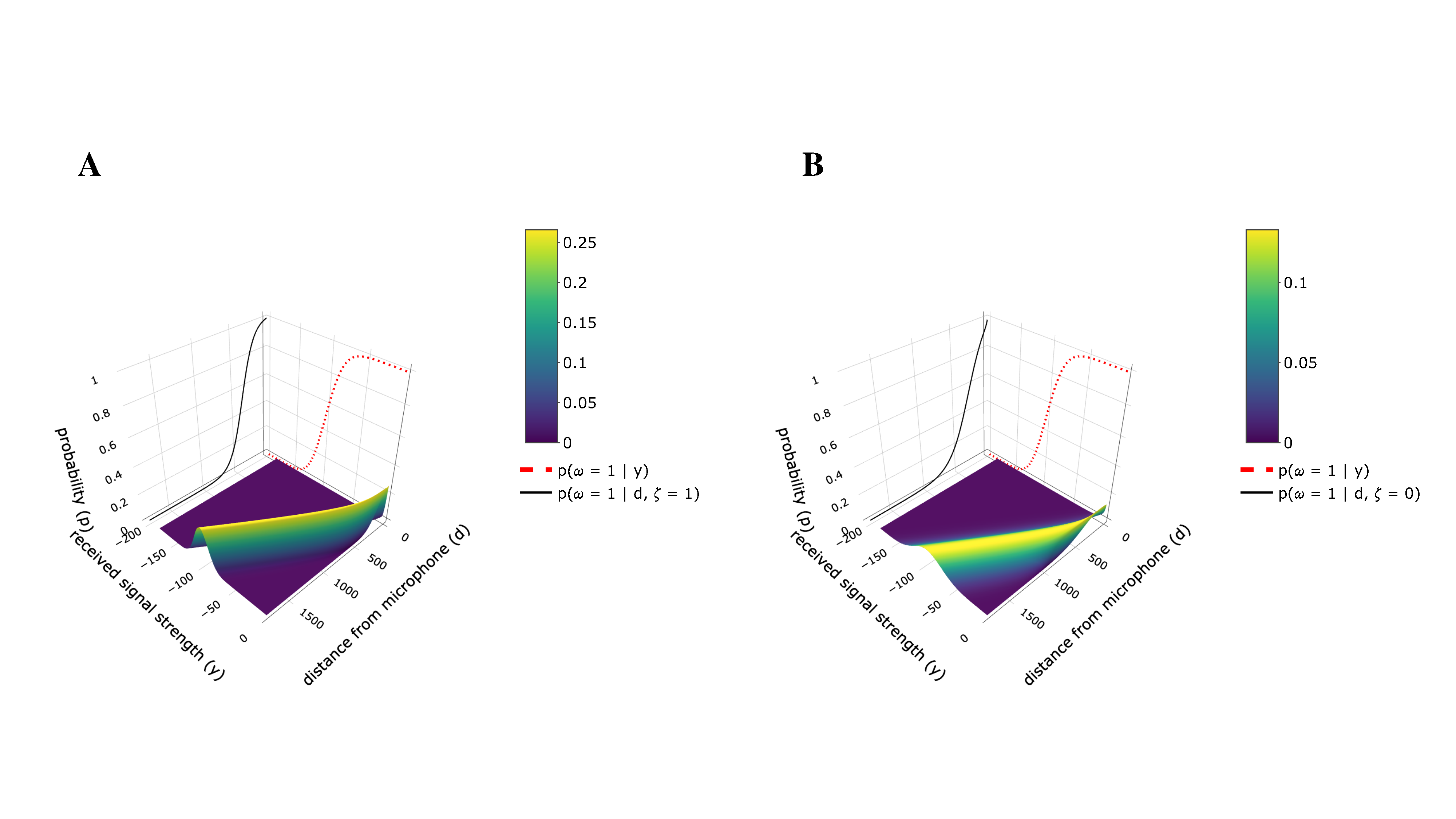}
          \caption{The detection function and its components for true positives (panel \textbf{A}) and false positives (panel \textbf{B}). See Figure~\ref{fig:compare_auto_step} for descriptions of the components. Note that the distributions of received signal strengths (the ``hills'' at the bases of the plots) are different for true positives and false positives. This figure appears in color in the electronic version of this article.}
         \label{fig:compare_tp_fp}
\end{figure}

The distance-dependent detection function for non-target calls is obtained in the same way as that for target-species calls shown in Figure \ref{fig:compare_tp_fp}\textbf{A} but with $f(y | d, \zeta = 0)$
instead of $f(y | d, \zeta = 1)$ at the base of the figure, resulting in a different detection function, $g(d, 0) = p(\omega = 1 | d, \zeta = 0)$, in the back left panel of Figure \ref{fig:compare_tp_fp}\textbf{B}:
\begin{equation} \label{equ:dp_fp}
    g(d, 0) = \int_{-\infty}^{\infty}p(\omega =1 | y)f(y | d, \zeta = 0)dy
\end{equation}

Like the pdf of $y$ for true positives, the pdf of $y$ for false positives, $f(y | d, \zeta = 0)$, is assumed to be Normal, but with parameters $\beta_0$ and $\sigma_s$ replaced by $\beta_0^{fp}$ and $\sigma_s^{fp}$. This is based on the assumption that identified calls that are not from target species may have a different mean and range of source signal strength. We assume that the rate of decay of these sounds governed by the parameter $\beta_1$ is the same as that of target species calls. The form of the decay function is also assumed to be the same for false positives and true positives \citep[see][and Web Appendix A for details in signal strength decay function]{https://doi.org/10.1111/j.1365-2664.2009.01731.x}.

The likelihood for false positive observations has the same structure as for true positives, but conditioning on $\zeta = 0$:
\begin{equation} \label{equ:ascrfp}
L_{fp}(\bm{\gamma}_{fp},\bm{\phi}) = \prod_{n=1}^{N} \int_{A} f(\bm{y}_n, \bm{z}_{n}, \bm{\omega}_n,\bm{x}_n | \zeta_n = 0)d\bm{x}
\end{equation}
\noindent
where $\bm{\gamma}_{fp}=(\beta_0^{fp}, \beta_1, \sigma_s^{fp}, r_0, r_1)$. 

\subsubsection{ML Confidence Measure} \label{parag:conf_out}
The automated detector, when applied to the acoustic recording data, outputs a measure of confidence $\rho$ that a detected call is from the target species. In this application, the measure is a positive real number in $(0, \infty)$ which we map monotonically onto the interval $(0,1]$ using an inverse logit function (see Web Appendix C for details). 

The resulting measure $\rho_{n,m}$ is only recorded if the call $n$ is detected by microphone $m$. If a call can be detected by multiple microphones, then we have more than one $\rho$. We assume that we can identify which detections on different microphones are from the same call. In the following, we use the average confidence measure across all microphones that have detected the call so that for the call $n$, our measure is
\begin{equation} 
\bar{\rho}_n = \frac{1}{J}\sum_{m:\omega_{n,m} = 1}\rho_{n,m}
\end{equation}
where $J = \sum_{m = 1}^{M}\omega_{n,m}$ is the number of microphones that detect the call $n$.

\subsubsection{Fixed-Confidence Mixture Model} \label{subsec:fixed_weight}
The key to developing mixture models that accommodate both the true and false positives is the conditional probability mass function for $\zeta_n$ given the confidence measure $\rho_{n,m}$. Because we do not know whether a call identified by the automated detector is a target species call or not, we model the pdf of the observed data as arising from a mixture of true positive and false positive pdfs with mixture weights for observation $n$ of $f(\zeta_{n}=1| \bar{\rho}_{n})$ and $f(\zeta_{n}=0| \bar{\rho}_{n})$.

For our fixed-confidence mixture model, we treat the $\bar{\rho}_n$s as probabilities such that $f(\zeta_{n}| \bar{\rho}_{n})$ is a Bernoulli distribution with parameter $\bar{\rho}_{n}$. The mixture likelihood is then defined as:
\begin{equation}
    L_{f}(\bm{\gamma}_{f},\bm{\phi}) =\prod_{n = 1}^{N} \sum_{\zeta_{n}} \int_{A} f(\bm{y}_n, \bm{z}_{n}, \bm{\omega}_n,\bm{x}_n| \zeta_{n}) f(\zeta_{n}| \bar{\rho}_{n}) d\bm{x}.
\end{equation}
\noindent
where $\bm{\gamma}_{f} =(\beta_0, \beta_0^{fp}, \beta_1, \sigma_s, \sigma_s^{fp}, r_0, r_1)$.

The density estimator of Eq~\ref{equ:HTL_estimator} requires $N$ to be the number of true positives, but we don't know this. If all the $\zeta_n$s were known, we could calculate $N$ as the sum of these values. But as we don't know them, we estimate $N$ by the expected value of this sum, conditional on the observations $\bm{y}_n, \bm{z}_{n}, \bm{\omega}_n$, and probabilities $\bar{\rho}_n$ ($n=1,\ldots,N$). The conditional pdf of $\zeta_n$ is 
\begin{equation} 
f(\zeta_{n} | \bm{y}_n, \bm{z}_{n}, \bm{\omega}_n,\bar{\rho}_n) 	= 
\frac{f(\bm{y}_n, \bm{z}_{n}, \bm{\omega}_n | \zeta_{n}) f(\zeta_{n}| \bar{\rho}_{n})}
{\sum_{\zeta_n=0}^1 f(\bm{y}_n, \bm{z}_{n}, \bm{\omega}_n | \zeta_{n}) f(\zeta_{n}| \bar{\rho}_{n})}
\end{equation}
\noindent
where $f(\bm{y}_n, \bm{z}_{n}, \bm{\omega}_n | \zeta_{n}) =\int_Af(\bm{y}_n, \bm{z}_{n}, \bm{\omega}_n,\bm{x}_n| \zeta_{n})\;d\bm{x}$. It follows that the conditional expectation of $\zeta_n$, given $\bm{y}_n, \bm{z}_{n}, \bm{\omega}_n, \bar{\rho}_{n}$, is $f(\zeta_n = 1 | \bm{y}_n, \bm{z}_{n}, \bm{\omega}_n,\bar{\rho}_n)$ and our estimator of $D_c$ becomes
\begin{equation} \label{equ:HTL_estimator_fixedwt}
    \hat{D_c} = \sum_{n=1}^N\frac{\hat{\rho}_n}{\hat{p}a T}
\end{equation}
\noindent
where $\hat{\rho}_n$ is $f(\zeta_n = 1 | \bm{y}_n, \bm{z}_{n}, \bm{\omega}_n,\bar{\rho}_n)$ evaluated at the maximum likelihood estimates of the parameters.

\subsubsection{Random-Confidence Mixture Model} \label{sec:random_weight}
The $\rho_{n,m}$ output by the ML identifier are generally not probabilities, just measures of confidence, so it may be better to develop a model and estimator that does not treat them as fixed probabilities. In this section, we treat them as observations of random variables whose distribution depends on the unknown latent variables $\zeta_n$. Specifically, we assume that $\rho_{n,m}$ is a draw from a pdf $f(\rho_{n,m} | \omega_{n,m}, \zeta_n)$ $(n=1,\ldots,N; m=1,\ldots,M)$ that has a parameter vector $\bm{\tau}_0$ for $\zeta=0$, and $\bm{\tau}_1$ for $\zeta=1$. The parameter vector $\bm{\tau} = (\bm{\tau}_0, \bm{\tau}_1)$ is estimated separately on an independent dataset. 
Assuming independence of the $\rho_{n,m}$s, we have 
\begin{equation} \label{equ:prob_rho}
  f(\bm{\rho}_n | \bm{\omega}_n , \zeta_n) = \prod_{m=1}^{M}f(\rho_{n,m} | \omega_{n,m}, \zeta_n) 
\end{equation}
\noindent
where $\bm{\rho}_n$ is a vector comprised of all the $\rho_{n,m}$s for all the microphones that have detected the call $n$.

In this case, the likelihood of our random-confidence mixture model is defined as follows:
\begin{longequation}
\begin{array}{ll}%
    L_{r}(\bm{\gamma}_{r},  \bm{\phi}, \pi) &= 
    \prod_{n = 1}^{N} \sum_{\zeta_{n}} \int_{A} \label{equ:mix_likeli} 
    f(\bm{y}_n, \bm{z}_{n}, \bm{\omega}_n,\bm{x}_n, \bm{\rho}_{n}|\zeta_{n}) 
    f(\zeta_{n}) d\bm{x} \\
    &= \prod_{n = 1}^{N} \sum_{\zeta_{n}} \int_{A} 
    f(\bm{y}_n, \bm{z}_n, \bm{\omega}_n,\bm{x}_n| \zeta_n) 
    f(\bm{\rho}_n |\bm{\omega}_n , \zeta_n) 
    f(\zeta_{n}) d\bm{x}
\end{array}
\end{longequation}
where $f(\zeta_{n})$ is a Bernoulli distribution with parameter $\pi$, which is the mixture weight in the likelihood and is the unconditional probability that an observation is a true positive. And the $\bm{\gamma}_r, \bm{\phi}$ have the same parameter components as the fixed-confidence mixture model.


We consider two models for $f(\rho_{n,m} | \omega_{n,m}, \zeta_n)$. We use a gamma distribution when $\rho$ is an output in the interval $(0,\infty)$, and a Beta distribution when $\rho$ is an output in the interval $(0,1]$. As is the case with received signal strength, we define $f(\rho_{n,m} | \omega_{n,m}, \zeta_n; \bm{\tau})$ to be 1 when $\omega_{n,m}= 0$ since $\rho$ is only recorded for detection.

Using arguments similar to those used for the fixed-confidence mixture model, the conditional expectation of $\zeta_n$, given $\bm{y}_n, \bm{z}_{n}, \bm{\omega}_n, \bm{\rho}_{n}$, can be shown to be 
\begin{equation} \label{equ:condi_expect}
f(\zeta_{n}=1 | \bm{y}_n, \bm{z}_{n}, \bm{\omega}_n,\bm{\rho}_n) 	= 
\frac{f(\bm{y}_n, \bm{z}_{n}, \bm{\omega}_n | \zeta_{n}=1) f(\bm{\rho}_{n} | \bm{\omega}_{n}, \zeta_{n}=1)f(\zeta_n = 1)}
{\sum_{\zeta_n=0}^1 f(\bm{y}_n, \bm{z}_{n}, \bm{\omega}_n | \zeta_{n}) f(\bm{\rho}_{n} | \bm{\omega}_{n}, \zeta_{n})f(\zeta_{n})}
\end{equation}
\noindent
and our estimator of $D_c$ becomes
\begin{equation} \label{equ:HTL_estimator_randwt}
    \hat{D_c} = \sum_{n=1}^N\frac{\hat{\pi}_n}{\hat{p}a T}
\end{equation}
\noindent
where $\hat{\pi}_n$ is $f(\zeta_n = 1 | \bm{y}_n, \bm{z}_{n}, \bm{\omega}_n,\bm{\rho}_n)$ evaluated at the maximum likelihood estimates of the parameters.

\subsubsection{Pseudo-likelihood Model} \label{subsec:weighted_data}
In addition to the mixture model, we propose to use observation confidence $\bar{\rho}$ as the power weight to calibrate the ASCR likelihood. The power weight can be seen as observing the $n$th capture history for $\bar{\rho}_{n}$ times \citep{10.1109/TPAMI.2016.2522425}. The intuition is that observations with low values of $\bar{\rho}_n$ will contribute less to the likelihood than those with high values. We consider this as the pseudo-likelihood, which is defined below: 
\begin{equation} 
L_{p}(\bm{\gamma}_{w},\bm{\phi}) =\prod_{n = 1}^{N} \left(\int_{A} f(\bm{y}_n, \bm{z}_n, \bm{\omega}_n,\bm{x}_n;\bm{\gamma})d\bm{x}\right)^{\bar{\rho}_{n}}
\end{equation}
As observations with low confidence are more likely to be false positives, incorporating weight into observation can effectively reduce bias in likelihood inference by mitigating the impact of such observations.

The parameter vector is $\bm{\gamma}_{p}=(\beta_0$, $\beta_1$, $\sigma_s$, $r_0$, $r_1)$. Because $\bar{\rho}_{n}$ is an unbiased estimator of $E[\zeta_{n}]$, we estimate $D_c$ by 
\begin{equation} \label{equ:HTL_estimator_wtdata}
    \hat{D_c} = \sum_{n=1}^N\frac{\bar{\rho}_{n}}{\hat{p}a T}
\end{equation}

\subsection{Bootstrap Procedure} \label{sec:bootstrap}
In the previous point estimation with likelihood, we make a simplified assumption on the independence of call locations, however, individuals can emit more than one call from the same location and we do not know which calls come from which individuals. The assumption will not affect the point estimation in general but have a substantial effect on interval estimation \citep{stevenson2015}. Therefore, we estimate the uncertainty of parameters and obtain interval estimates using a parametric bootstrap following \citet{stevenson2015}.

In order to speed up the bootstrap procedure, we use a rejection sampling process \citep{10.2307/4356322}. That is, for each received signal strength, we use our estimator of $p(\omega = 1 | y)$ parameterised with $\hat{r}_0$ and $\hat{r}_1$ to obtain the probability of success detection, with which we sample from a Bernoulli distribution to determine whether a signal is detected. 

We train an ML model on the training data and then apply the model on an independent labelled dataset (usually named validation set) to obtain the confidence output on each sample in the validation set, thus allowing us to sample the confidence for true positives and false positives separately. 

The animal density $D_a$ is set to be $D_c / \mu_c$ where $\mu_c$ is the mean animal call rate. We only focus on the estimation of $D_c$ in this application while the way of estimating $D_a$ can be easily integrated into our model with the method proposed in \citet{stevenson2015}.

In the following, we describe the bootstrap procedure for the random-confidence mixture model. Bootstrap procedures for the other models are similar. The simulated data or parameters estimated from simulated data are denoted with the superscript $*$: 


\begin{enumerate}
\item Simulate animal location as a realization of a homogeneous Poisson process with intensity $\hat{D}_{a}$.

\item Generate $\bm{X}^{*}_{tp}$ by repeating each location from Step(1) $\mu_c$ times, where $\mu_c$ is the constant call rate. 


\item Sample $\bm{P}^*_{tp}$ from true positives from the validation set.

\item Obtain $\bm{\Omega}^*_{tp}$ by simulating from the estimate of $f(\omega_{n,m} | \bm{x}_{n}^{*}, \zeta_n = 1 )$ with 
Eq \ref{equ:dp_auto} using rejection sampling. 

\item Obtain $\bm{Y}^{*}_{tp}$ by simulating from the estimate of $f(y_{n,m}|\omega_{n,m}^{*} = 1, \bm{x}_{n}^{*}, \zeta_n = 1)$ with Eq \ref{equ:ss_auto} and $\bm{Z}^*_{tp}$ by simulating from the estimate of $f(\bm{z}_{n}|\bm{\omega}_{n}^{*}, \bm{x}_{n}^{*})$ with Eq \ref{equ:toa} (see Appendix B
for details) for all observations.

\item Calculate the false positive rate $\hat{f} = 1 - \sum_{n=1}^{N}\frac{\hat{\pi}_{n}}{N}$ using the conditional expectation of $\zeta_{n}$s with Eq \ref{equ:condi_expect}. 

\item Set the noise observation number $N_{fp}$ with false positive rate $\hat{f}$, and the true positive number $N_{tp}$ generated in the above steps.

\item Simulate noise location $\bm{X}^{*}_{fp}$ as independent Uniform distribution in a survey area $a$ for $N_{fp}$ times. 

\item Sample $\bm{P}^*_{fp}$ from false positives from the validation set.

\item Obtain $\bm{\Omega}^*_{fp}$,  $\bm{Y}^*_{fp}$, $\bm{Z}^*_{fp}$ using the same procedure in Steps 4-5, while simulating from $f(\omega_{n,m} | \bm{x}_{n}^{*}, \zeta = 0)$, $f(y_{n,m}|\omega_{n,m}^{*} = 1, \bm{x}_{n}^{*}, \zeta = 0)$ and $f(\bm{z}_{n}|\bm{\omega}_{n}^{*}, \bm{x}_{n}^{*})$ used in the likelihood \ref{equ:ascrfp}.

\item Generate $\bm{P}^* = \{\bm{P}^*_{tp}, \bm{P}^*_{fp} \}$,
$\bm{\Omega}^* = \{\bm{\Omega}^*_{tp}, \bm{\Omega}^*_{fp}\}$,  $\bm{Y}^* = \{\bm{Y}^*_{tp}, \bm{Y}^*_{fp}\}$, $\bm{Z}^* = \{\bm{Z}^*_{tp}, \bm{Z}^*_{fp}\}$ by combining true positives and false positives. 

\item Calculate $\hat{\bm{\gamma}}^{*}, \hat{\bm{\phi}}^{*}, \hat{\pi}^{*}$ from $\bm{\Omega}^*$, $\bm{Y}^*$, $\bm{P}^*$,and $\bm{Z}^*$ using likelihood \ref{equ:mix_likeli} and $\hat{\bm{\tau}}$ estimated from the validation set.

\item Calculate $\hat{D}^*_c$ with Eq \ref{equ:HTL_estimator_randwt} and $\hat{D}^*_a = \hat{D}^*_c / \mu_c$.

\item Repeat the above steps $B$ times and save the parameter estimates from each iteration.


\end{enumerate}

\section{Simulation} \label{sec:simulation}
The use of digital recorders in ASCR surveys is new and we are not aware of any such dataset with adequate data to train an ML detector and provide data adequate for our method. Because the update of digital ASCR is increasing, we expect that such datasets will soon be available and present our method as a means of doing inference with them when they are available. Meanwhile, we evaluate our methods using a simulation study. We try to make the simulation as close to reality as possible by using recordings of Hainan gibbon ($\it{Nomascus \ hainanus}$) calls from \citet{Dufourq2020.09.07.285502}. However, the data were gathered with microphones separated by too great a distance to be detections of any call on more than one microphone, so they are not directly amenable to ASCR analysis. 

 In the survey, the source signal strength $\beta_0$ is set to 0 (dBFS), and the linear decay of signal strength $\beta_1$ is set to 0.12. The parameter $\sigma_{t}$ in $\bm{\phi}$ (see Appendix B
 for details) is set as 2, which controls Gaussian measurement error for the time of arrival. The recorded signal strength standard deviation $\sigma_{s}$ is set to 15, similar to the value used in \citet{stevenson2015}. With this parameter, the received signal strength distribution at each microphone is similar to that in the audio recording dataset. 

Corresponding to the signal strength parameters set above, we set 16 detectors $\{M_m | m = (1,2, ..., 16)\}$ separated by 600m in both X-direction and Y-direction, and the minimum distance between detector locations and the edge of the generated survey area is set to 1800m. The population call density $D_c$ is set as 0.06 per hectare and the call rate $\mu_c$ is a scalar value of 0.5 calls per hour. The reason is that the real density of gibbons is around 0.04 to 0.21 per hectare, and the Hainan gibbon is one of the rarest among them. The survey duration $T$ is set to 8 hours. We set the simulation density at a very low value in order to test the model's capacity to deal with small sample sizes. 

We assume that false positive sound sources have lower mean source signal strength $\beta^{fp}_0 = -15$ and greater range $\sigma^{fp}_s = 30$ since false positives can come from various sources. We set the signal strength decay rate for false positive $\beta^{fp}_1 = 0.12$ the same value as for true positive under the assumption that the false positive and true positive signals are propagated in the same way when in the same environment. We control the number of false positive observations with the false positive rates $f$ in the detection model that is evaluated on the test data. We do the simulation in two steps: in the first, we do not add false positives while in the second step, we add false positives to the simulated dataset. 

\subsection{Data Description}

The dataset contains 25 8-hour recordings of Hainan gibbon calls collected in Bawangling National Nature Reserve, Hainan, China, with eight Song Meter SM3 recorders. Recordings last eight hours each day, with an acoustic sampling rate of 9.6KHz and a bit depth of 16. The dataset contains a total of 1,858 gibbon calls in 9,199 seconds. 

We split the whole dataset into training, validation, and test set.  The training set is used for the ML model training procedure. The validation set is used to estimate the parameters $\hat{\bm{\tau}}$. During the bootstrap procedure, we sample confidence value $P_{val}$ for true positive and false positive observation separately from the validation set. The test set is used for obtaining value for detection probability parameters $r_0, r_1$ and sampling the confidence value $P_{test}$ along with false positive rate $f$. We then use this information to simulate the capture histories. 

\subsection{Detection model}
We apply a convolutional recurrent neural network introduced by \citet{https://doi.org/10.1111/2041-210X.13873} for automated gibbon call detection. The ML model is applied to the test data and then the logistic regression is fitted with received signal strength and detection states, indicating whether a call is detected or not. Then we obtain the mean and variance for $r_0$ and $r_1$, assuming to be asymptotic Normal, from which we sample $r_0$ and $r_1$. Then we use rejection sampling with parameters $r_0, r_1$ to generate simulation data. More specifically, for point estimation, we randomly sample 1000 detection parameters $r_0, r_1$, confidence set $P$, and false positive rate $f$ from the detection model applied to the test set.
We then generate 1000 datasets using the pre-set parameters (e.g. $\beta_0$, $\beta_0^{fp}$, $\beta_1$, $\beta_1^{fp}$, $\sigma_s$, $\sigma_s^{fp}$, $\sigma_t$, $D_c$,) along with sampled $r_0, r_1, P, f$. 
For the bootstrap, we generate 200 simulated datasets in the same way as above.

\subsection{Results}
In this section, we compare the performance of the proposed model via point estimation and interval estimation of the simulated call density. We then compare the average time consumption of different models and the GPU-accelerated models. 

\subsubsection{Point Estimation}

\begin{figure}[]   
         \centering
         \includegraphics[width=7in]{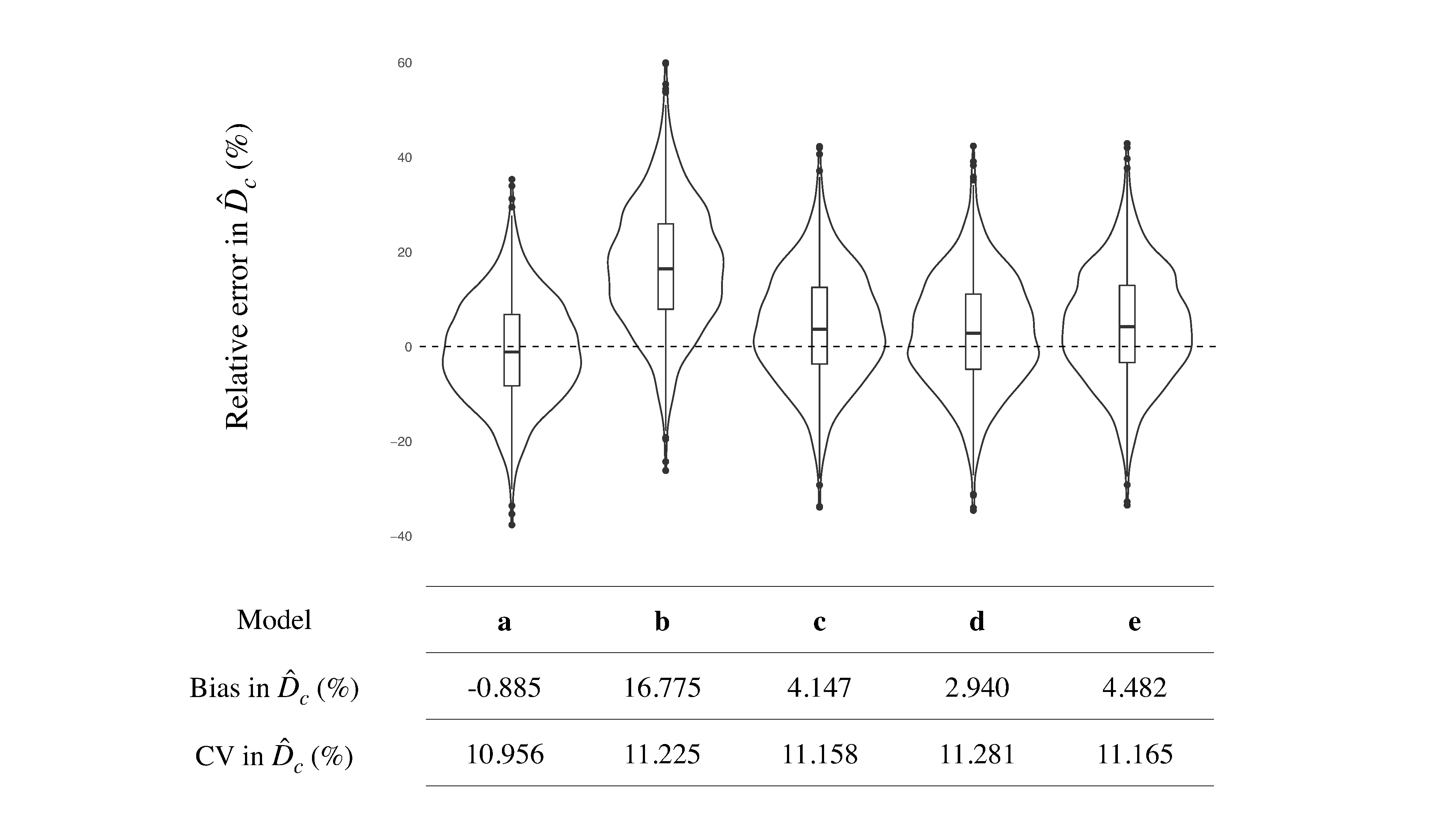}
         \caption{Relative error, bias and CV of point estimation of the following models: \textbf{a:} Automated ASCR (without false positive observations). \textbf{b:} Automated ASCR (with false positive observations). \textbf{c:} Fixed-confidence mixture model. \textbf{d:} Random-confidence mixture model. \textbf{e:} Pseudo-likelihood model. The models \textbf{c, d, e} all have false positives added to the observations. The simulated data are generated with $D_c$ as 0.06. 
         }
         \label{fig:point_estimate}
\end{figure}

In point estimation, bias is calculated as a percentage of $\hat{E}(\hat{D}_c - D_c)$ to $D_c$ and the expectation is acquired by taking the average over 1000 simulation results. We also calculate the coefficient of variation (CV) as a percentage of the true value in point estimation. 
We compare the point estimation performance of the proposed methods in Figure \ref{fig:point_estimate}. 



The automated ASCR using false positive-free data achieves negligible absolute bias of less than 1\%; i.e., -0.89\% of the estimated size, while it produces nearly 17\% bias when adding false positives to the dataset. The mixture models all produce less than 5\% bias, among which the random-confidence mixture model has the lowest bias; i.e., only 2.94\%. The pseudo-likelihood model also produces less than 5\% bias. All methods produce a similar CV. 

\subsubsection{Interval Estimation}

We apply bootstrap confidence interval methods named \textit{normal}, and \textit{percentile} to estimate the confidence interval of parameters. The \textit{normal} method assumes the parameter follows a Normal distribution, while the \textit{percentile} methods use quantile limits for interval estimation \citep{davison_hinkley_1997}. 
The interval estimation performance of the models is shown in Table \ref{tab:interval_estimation}.

\begin{table}
\caption{Coverage of \textit{normal}, and \textit{percentile} confidence interval methods for the parameter $D_c$ estimated by proposed methods. Nominal coverage is set as 95\%. All confidence interval methods rely on the bootstrap procedure. }
\label{tab:interval_estimation}
\begin{center}
\begin{tabular}{lcc}
\Hline
Estimator                                            & \textit{Normal} & \textit{Percentile} \\ \hline
Automated ASCR (without false positive observation)    & 0.965  & 0.945      \\
Automated ASCR (with false positive observation)    & 0.715  & 0.780      \\
Fixed-confidence mixture model                         & 0.930  & 0.950      \\
Random-confidence mixture model  & 0.930   & 0.955     \\ 
Pseudo-likelihood model                          & 0.925  & 0.950     \\
\hline
\end{tabular}
\end{center}
\end{table}

When simulating data without false positive observations, the automated ASCR method yields coverage rates similar to the nominal 95\% rate with both confidence interval calculation methods. In simulations that include false positive observations, the \textit{normal}, and \textit{percentile} confidence interval methods produce poor coverage rates of only 0.715 and 0.78 respectively. 

In contrast, all models that are designed for dealing with false positives achieve coverage rates similar to the nominal 95\% rate with all confidence interval calculation methods. Among them, the pseudo-likelihood model improves the coverage rate to 0.925, and 0.950 with the \textit{normal} and \textit{percentile} confidence interval methods respectively. The random-confidence mixture model yields a coverage rate of 0.955, and the fixed-confidence method achieves a coverage rate of 0.950 using the \textit{percentile} method.


It is worth noting that the pseudo-likelihood model does not have the ability to bootstrap intuitively since it does not model the false positive signal parameters (i.e. $\beta^{fp}_0, \sigma^{fp}_s$). But we can achieve interval estimation by modelling the false positive data alone with an independent labelled dataset, where we treat all false positive observations as from one target species and apply automated ASCR to model them. 

\subsubsection{Computation Cost}
We compare different models' run times in Table \ref{tab:time_consumpt}. The results are averaged over 1000 repetitions. The automated ASCR model and pseudo-likelihood model have similar run times of about 250 seconds while the mixture model takes nearly three times as long to fit. After rewriting code using matrix operations to make it suitable for use with a GPU, we were able to reduce the mixture model's run time to less than 21s, and the pseudo-likelihood model's run time to only 6.54s. 


\begin{table}
\caption{The average time consumption in seconds for parameter inference over 1000 repetitions of automated ASCR model, fixed-confidence mixture model, random-confidence mixture model, pseudo-likelihood model and all four models with GPU acceleration.}
\label{tab:time_consumpt}
\begin{center}
\begin{tabular}{lc}
\Hline
Estimator                                             & Average time consumption (s) \\ \hline
Automated ASCR                                        & 243.16  \\
Fixed-confidence mixture model                       & 641.98   \\
Random-confidence mixture model                   & 631.23 \\ 
Pseudo-likelihood model                             & 262.59  \\
Automated ASCR + GPU                                  & 5.81  \\
Fixed-confidence mixture model  + GPU                  & 20.45  \\
Random-confidence mixture model  + GPU                  & 13.44  \\
Pseudo-likelihood model + GPU                          & 6.54  \\
\hline
\end{tabular}
\end{center}
\end{table}

\section{Discussion}

\subsection{Comparison to Canonical Estimators}
There are two key differences between our method and the widely used canonical estimator method of \cite{https://doi.org/10.1111/brv.12001} given by Eq~(\ref{equ:canonical}). Firstly, unlike the canonical method, our method takes account of the existence of false positives in estimating the mean detection probability $p$. Secondly, it uses the information in the observations associated with each detected call to estimate the expected probability that a call is a true positive ($E[\zeta_n]$ for the $n$th call), whereas the canonical estimator uses the uniform expected probability for all the detected calls, irrespective of the ML confidence measure or any other observed data associated with the call. More specifically, the canonical estimator of call density can be written as 
\begin{eqnarray}
\hat{D}_c&=&\sum_{n=1}^N\frac{(1-\hat{f})}{\hat{p}aT}
\end{eqnarray}
\noindent
where $(1-\hat{f})$ is an estimate of the probability that an observation is a true positive. In contrast, our estimators have the form
\begin{eqnarray}
\hat{D}_c&=&\sum_{n=1}^N\frac{\hat{E}[\zeta_n]}{\hat{p}aT}
\end{eqnarray}
\noindent
where $\hat{E}[\zeta_n]$ is an estimate of the probability that the detected call $n$ is a true positive, from all the information associated with the call, including the spatial information and the ML measure of confidence that it is a call from the target species.

Our estimators are able to discriminate between calls that are more likely to be true positives and those that are less likely to be true positives on the basis of the observations associated with them. However, the canonical estimator cannot do this. Moreover, our estimator uses the unlabelled data observed on the survey itself in addition to the ML model trained on labelled data, to estimate this probability, whereas the canonical estimator relies entirely on the data used to train the ML model and not the unlabelled \textit{in situ} from the survey. And the \textit{in situ} mean probability of false positives may not be the same as that in the data used to train the ML model.

Our method also allows for the fact that the data used to estimate the mean detection probability $\hat{p}$ may contain false positives, whereas the canonical method implicitly assumes that it does not, or at least that it is not affected by the presence of false positives. In short, our methods are more flexible and versatile, and if the false positive probability is actually the same for all observations, and false positives do not affect $\hat{p}$, then our estimator reduces to the canonical estimator.

\subsection{Comparison among Proposed Methods}
 When using ML detection output that has no false positives, ASCR performs well, but both the point and interval estimates may be considerably biased in the presence of false positives. 
 
The pseudo-likelihood model is the simplest and fastest among the methods we propose for dealing with false positives. In our simulations, this model reduced bias from 17\% to just 4\% and has a coverage probability close to the nominal 95\%. However, it requires modelling $\bm{\gamma}_{fp}$ using a separate dataset for the bootstrap procedure. This is because it does not estimate these parameters. This requires a labelled dataset that shares the same false positive parameters as the survey data.
 
The mixture model performs best, with negligible bias and coverage probability very close to the nominal value. This comes at the cost of nearly doubling computing time and memory requirements, but this cost is very small in comparison to the time and effort required to do an ASCR survey. 

The key ideas underpinning the ASCR mixture model are to use the measure of confidence from the ML detector as a covariate related to true/false positive status, and to treat true/false positive status as a latent variable. These ideas are applicable to other survey methods, like non-acoustic spatial capture-recapture, capture-recapture, distance sampling and occupancy methods, whenever ML is used for object identification.









\backmatter


\section*{Acknowledgements}

The authors thank Dr Ben Stevenson for helpful suggestions. YW is partly funded by the China Scholarship Council (CSC) for Ph.D. study at the University of St Andrews, UK. \vspace*{-8pt}

\section*{Data Availability Statement}

The data that support the findings of this study are openly available in Zendo at \\ http://doi.org/10.5281/zenodo.3991714, reference number 3991714. 


%
 \bibliographystyle{biom} 
\bibliography{autodetscr}






\section*{Supporting Information}

Web Appendix A, B, and C referenced in Section~\ref{sec:method}, and Web Appendix D for software implementation are available with
this paper at the Biometrics website on Wiley Online
Library. 
\vspace*{-8pt}

\appendix


\section{A} 

\subsection*{Approximation to Sigmoid-Normal integration} \label{Appen:fast_integrate}

Note that the likelihoods \ref{equ:dp_auto} and \ref{equ:ss_auto} require the calculation of the integral over products of a gradual function and a Gaussian distribution, while the numerical integration can be time-consuming or inaccurate. Following \citet{https://doi.org/10.48550/arxiv.1703.00091}, we employ an approximation method to the expectation of Sigmoid function over Normal distribution; that is:
\begin{equation}
    \int Sigmoid (r_1  y + r_0) N(y|\mu, \sigma^2) dy \approx  Sigmoid(\frac{r_1 \mu + r_0}{\sqrt{1 + \lambda(r_1 \sigma)^2}})
\end{equation}
where we have $\hat{\lambda} = 0.368$ based on Monte Carlo estimation.

\section{B} 
\subsection*{The pdf of Received Signal Strength} \label{Appen:pdf_ss}
Since signal strength $y$ is only recorded if a call is detected, we model the observed signal strength conditional on detection ($\omega=1)$ using the Bayes' rule:
\begin{equation} \label{equ:ss_auto}
    f_y(y | d, \zeta = 1, \omega = 1) = \frac{p(\omega = 1 | y) f_y(y | d, \zeta =1 )}
    {\int_{-\infty}^{\infty} p(\omega = 1 | y) f_y(y | d, \zeta =1 )dy}
\end{equation}
\noindent

We assume the signal strength observations are independent given latent location $\bm{x}_n$:

\begin{equation} \label{equ:ss_likelihood}
   f(\bm{y}_n|\bm{\omega}_n,\bm{x}_n, \zeta_n =1) = \prod_{m = 1}^{M}f(y_{m,n}|\omega_{m,n}, \bm{x}_n, \zeta_n = 1)
\end{equation}
where $f(y_{m,n}|\omega_{m,n}, \bm{x}_n, \zeta_n = 1)$ is defined in equation \ref{equ:ss_auto}.

\subsection*{The pdf of Detection Time} \label{Appen:pdf_toa}

The time of arrival likelihood remains unchanged as the standard ASCR model \citep{stevenson2015}. When accounting for uncertainty in record time due to Gaussian measurement error, which is controlled by the parameter $\sigma_t$, we can write the density function for the time of arrival difference as:
\begin{equation} \label{equ:toa}
    f(\bm{z_n}|\bm{\omega}_n, \bm{x}_n) = \frac{(2\pi \sigma_t^2)^{(1-J_n)/2}}{T\sqrt{J_n}}exp(\sum_{m: \omega_{n,m} = 1}\frac{(\delta_{n,m}(x_n) - \bar{\delta_n})^2}{-2\sigma_t^2})
\end{equation}

where $J_n$ is the number of microphone that detected call $n$; that is, $J_n = \sum_{m = 1}^{M}\omega_{n,m}$. And $\delta_{n,m}(x_n) = z_{n,m} - d_m(\bm{x}_n)/v$ is expected call production time, in which $v$ is speed of sound and $d_m(\bm{x}_n)$ is the distance between the location of call $n$ and  detector $m$. $\bar{\delta}_n$ is the average production time for the call $n$ across all detectors. When $J_n = 1$, we set the  $f(\bm{z_n}|\bm{\omega}_n, \bm{x}_n) = 1$.

\subsection*{The pdf of Capture History} \label{Appen:pdf_bincapt}
We assume the detection between $M$ microphones to be independent conditioning on the call latent location $\bm{x}_n$. Based on the detection function, the probability of detection likelihood for one observation across $M$ detectors is:
\begin{equation} \label{equ:det_likelihood}
    f(\bm{\omega}_n|\bm{x}_n, \zeta_n =1) = 
    \frac{\prod_{m = 1}^{M}f(\omega_{m,n}|\bm{x}_n, \zeta_n = 1)}{p.(\bm{x}_n| \zeta_n = 1)}
\end{equation}
where this probability is conditioned on a call being detected by at least one microphone:
\begin{equation} 
    p.(\bm{x}_n| \zeta_n = 1) = 1 - \prod_{m=1}^{M} 1 - g(d_{m}(\bm{x}_n) , 1)
\end{equation}
and $f(\omega_{m,n}|\bm{x}_n, \zeta_n = 1)$ is a Bernoulli random variable with $g(d_{m}(\bm{x}_n) , 1)$ as parameter.

\subsection*{The pdf of the Source Location} \label{Appen:pdf_loc}
If we assume call locations to be all independent, then the call location $\bm{x}_n$ is a realization of a filtered homogeneous Poisson point process:
\begin{equation}
    f(\bm{x}_n | \zeta_n = 1) = \frac{p.(\bm{x}_n |\zeta_n = 1)}{\int_{-\infty}^{\infty} p.(\bm{x}_n |\zeta_n = 1)d\bm{x}}
\end{equation}
and we have defined the probability that a call has been detected at least once $p.(\bm{x}_n |\zeta_n = 1)$ above.

\section*{Web Appendix A}
We use the following signal strength decay function:
\begin{longequation}
E[y|d; \beta_0, \beta_1] =
\left\{\begin{array}{cc}%
\beta_0 -  20 \times log_{10}(d) - \beta_1 \times (d-1) & \ d > 1 \\ \beta_0 & \quad d<=1
\end{array}
\right.
\label{equ:decay_fp}
\end{longequation}
where $\beta_0$ is the source signal strength, $\beta_1$ is the linear decay of the signal strength, and $d$ is the distance between the signal source and the detector. 
This is the same as the signal strength decay ``full model'' used in \citet{https://doi.org/10.1111/j.1365-2664.2009.01731.x}, which produces the most reliable estimation results for all parameters compared to other attenuation models.

\section*{Web Appendix B}
Different ML detection models may result in varied detection probability curve shapes (i.e. S-shape curve in our case). The functional form of the detection probability function should be decided with an independent separate dataset named validation set. We use the loudness of the denoised sound signal to represent the signal strength during this application. 

In our application, signal strength is calculated by using the root mean square (RMS) of the signal amplitude after denoising. Denoising is done using spectral gating \citep{spectral_gating}. 
This works by using pure noise data in the vicinity of a detected target call to estimate a noise threshold value for each frequency band. This estimated threshold is then used to mask the target audio clip, meaning that any sounds in the target audio that fall below the estimated threshold for a given frequency band are silenced or suppressed. By using this noise threshold masking approach, the method is able to better isolate and extract the desired signal from noisy audio recordings.

By taking the detection state (binary variable indicating call is detected or not) as the dependent variable and denoised signal strength as the independent variable, we apply logistic regression to the validation set, as shown in Web Figure 4. 

\renewcommand{\figurename}{\textbf{Web Figure}}

 \begin{figure}[]
	\centering
	\includegraphics[width=6in]{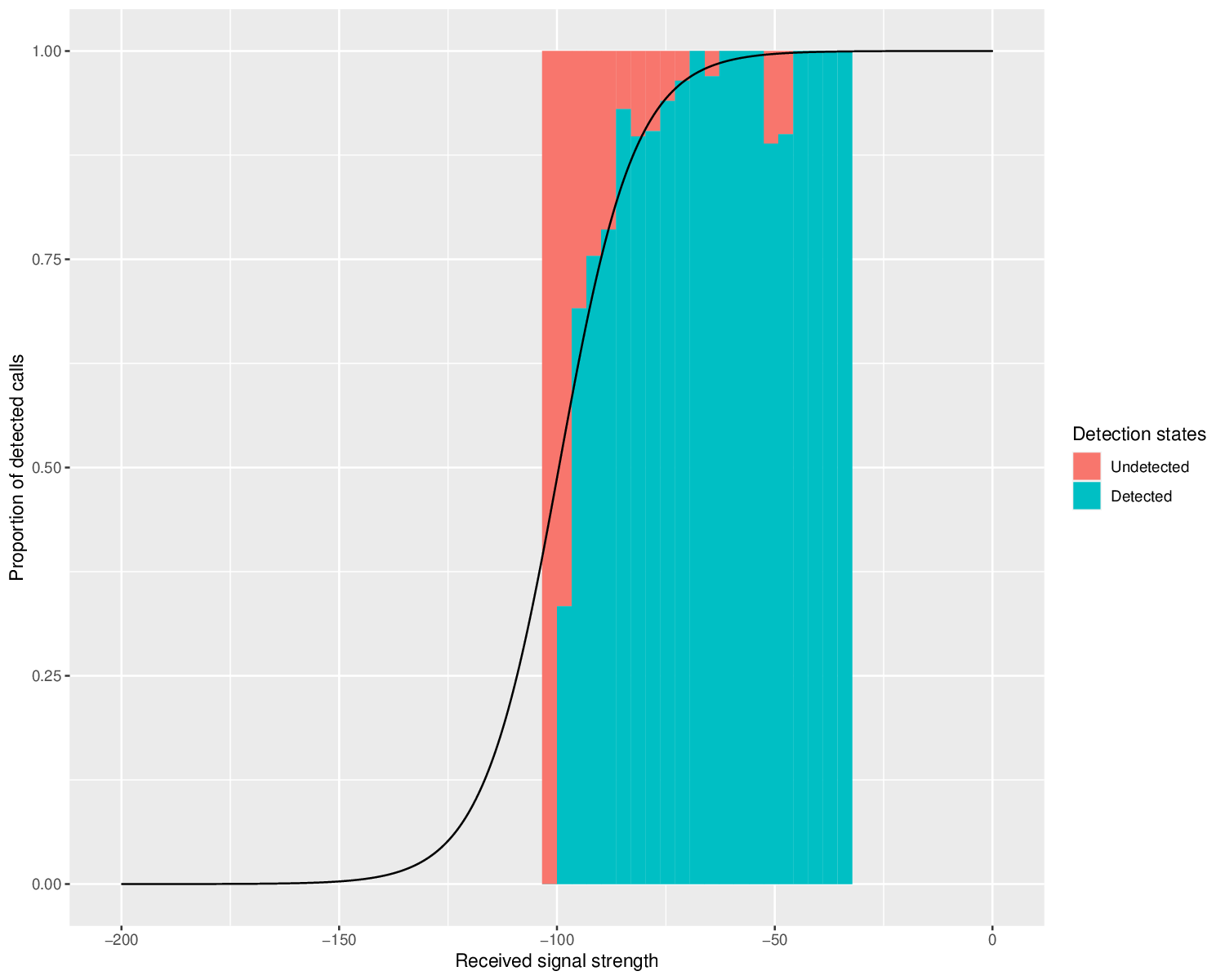} 
	\caption*{\textbf{Web Figure 1:} The form of detection probability depends on the signal strength and the logistic regression result (black curve); the x-axis represents signal strength and the y-axis represents the proportion of calls detected ($1 - false \ negative \ rate$). This figure appears in color in the electronic version of this article, and color refers to that version.} 
    \label{fig:logistic_fit}
\end{figure}
The logistic regression provides us with the detection function according to the denoised signal strength. We then have the continuous detection function with the Sigmoid (logistic) functional form:
\begin{equation} 
f(y) = Sigmoid(r_1 y + r_0) = \frac{1}{ 1+ e^{-(r_1 y + r_0)}}
\end{equation}
Note that $\hat{r}_0, \hat{r}_1$ here are not necessarily used in automated ASCR inference as $f(y)$ is only used for determining the curve functional form.

\section*{Web Appendix C}
The raw output from the ML detection model is a positive real number since only the prediction with a positive value is counted as a detection; otherwise, ignored. We can then map this value onto $(0,1)$ space with the standard Sigmoid function: $Sigmoid(\rho) = \frac{1}{1 + e^{-\rho}}$, which is a common procedure in machine learning. However, according to \citet{pmlr-v70-guo17a}, modern networks, especially with negative log-likelihood (NLL) loss, tend to output poorly mapping confidence because the neural network will gradually overfit to NLL loss without overfitting to binary classification loss. Following \citet{Platt_scaling}, we calibrate the mapped confidence with parameter estimated from logistic regression: 
\begin{equation}
    f(\rho) = \frac{1}{1 + e^{-(\hat{a} \rho + \hat{b})}}
\end{equation}
where $\hat{a}, \hat{b}$ are estimated from the validation set by treating the true positive or false positive state as the dependent variable, and raw confidence output as the independent variable. According to \citep{pmlr-v70-guo17a}, this method is proven to be efficient in calibrating the CNN outputs. 

\renewcommand{\figurename}{\textbf{Web Figure}}

 \begin{figure}[]
	\centering
	\includegraphics[width=6in]{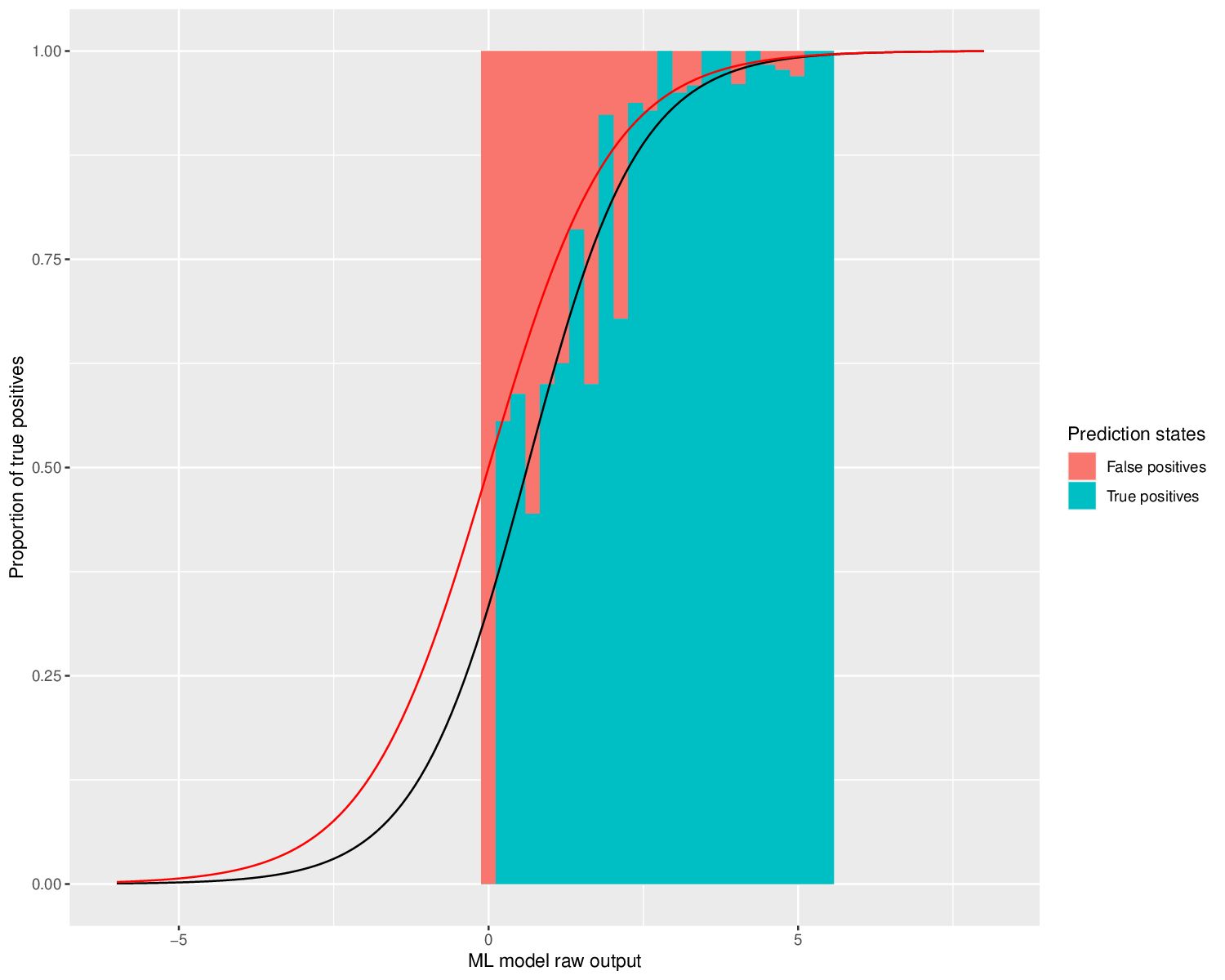} 
        \label{fig:confidence_cali}
	\caption*{\textbf{Web Figure 2:} The y-axis is the proportion of detected calls belonging to true positive detection and the x-axis is the ML model's raw output. The calibrated confidence is shown in a black line and un-calibrated confidence is shown in a red line. This figure appears in color in the electronic version of this article, and color refers to that version.} 
\end{figure}

As shown in Web Figure 5, the un-calibrated confidence (demonstrated in a red curve) tends to overestimate the confidence that a detection belongs to true positives, while calibrated confidence curve (demonstrated in a black curve) can mitigate this effect by refitting the ML raw output $\rho$ with observations' real state (This figure appears in color in the electronic version of this article, and color refers to that version). In this work we assume the false positive rate to be independent of signal strength level, thus the confidence does not have an effect on the call detection function.

\section{Web Appendix D}
All models are implemented using R with packages \textit{ascr}~\citep{stevenson2015ascr}, \textit{TMB}~\citep{10.18637/jss.v070.i05} and \textit{nlminb}, where \textit{ascr} is used for generating simulation data, \textit{TMB} for automated differentiation, and \textit{nlminb} for the optimization of the log-likelihood. We have applied the proposed algorithms in \textit{Pytorch} \citep{NEURIPS20199015} with GPU CUDA acceleration and managed to speed up the inference by a factor of about 30 times. 

\label{lastpage}

\end{document}